\documentclass{article}

     \usepackage[final]{neurips_2019}

\usepackage[utf8]{inputenc} 
\usepackage[T1]{fontenc}    
\usepackage{hyperref}       
\usepackage{url}            
\usepackage{booktabs}       
\usepackage{amsfonts}       
\usepackage{nicefrac}       
\usepackage{microtype}      

\newcommand{\be}{\begin{equation}}
\newcommand{\ee}{\end{equation}}

\title{Collecting Charges for Ad Impact on User Experience for Different Price Types}

\author{%
  Eric Bax\\
  \textit{ebax@verizonmedia.com}\\
}

\begin{document}

\maketitle

\section{Summary}
This note describes how to collect charges for ad impact on user experience. The charge may be per-view, to account for impact on user experience from viewing an ad, or per-click, to account for impact from clicking on the ad. The results for per-click charges also apply to per-conversion charges or per-action charges. Conceivably, a marketplace could assess both kinds of charges.

Different advertisers may use different price types. CPM advertisers agree to pay for each view. CPC advertisers agree to pay only in case of a click. The paper “Ad Auction Design and User Experience” \citep{abrams07} explains how to collect per-click charges from CPC advertisers. This brief note explains how to extend their result to collect per-view charges from CPC advertisers and per-view and per-click charges from CPM advertisers.

The main purpose of this note is to extend the Abrams and Schwarz work for the search advertising marketplace to the display advertising marketplace. In search advertising, the pricing is generally CPC, and the main impact on user experience results from clicking on an ad. Display advertising has a diversity of price types, including CPM, CPC, and CPA. Also, in display advertising, simply viewing an ad can have a strong impact on user experience.

\section{Collecting Per-View Charges for CPC Ads}
Let $v$ be the per-view charge for impact on user experience for an ad. Let b be the advertiser’s per-click bid. Let $p$ be the click probability for the ad in the position being auctioned. Then:
\begin{itemize}
\item Enter value $a = bp - v$ into the auction for the ad. (Enter $a$ as the per-view bid. Equivalently, enter $a/p$ as the per-click bid and $p$ as the click probability.)
\item If the ad wins and receives a click, charge $r + \frac{v}{p}$, where $r$ is the auction price of the click. (In a first-price auction, $r = a/p$. In a second-price auction, $r \leq a/p$.) 
\end{itemize}
Note that the full click charge $r + \frac{v}{p}$ is at most $b$.

\section{Collecting Per-View Charges for CPM Ads}
As in the previous section, let $v$ be the per-view charge. Let $b$ be the per-view bid. Then:
\begin{itemize}
\item Enter value $a = b - v$ into the auction for the ad. (Enter $a$ as the per-view bid.) 
\item If the ad wins, charge $r+v$, where $r$ is the auction price of the view. (In a first-price auction, $r = a$. In a second-price auction, $r \leq a$.)
\end{itemize}

\section{Collecting Per-Click Charges for CPM Ads}
Let $c$ be the per-click charge for ad impact on user experience. Let $b$ be the per-view bid. Let $p$ be the click probability for the ad in the position being auctioned. Then: \begin{itemize}
\item Enter value $a = b - cp$ into the auction for the ad.
\item If the ad wins, charge $r + cp$, where $r$ is the auction price of the view.
\end{itemize}

\section{Collecting Per-View and Per-Click Charges for CPC Ads}
Let $v$ be the per-view charge and $c$ be the per-click charge. Let $b$ be the per-click bid and $p$ be the click probability.
\begin{itemize}
\item Enter value $a = (b - c)p - v$ into the auction as the per-view bid for the ad. (Equivalently, enter $a/p$ as the per-click bid and enter $p$ as the click probability.)
\item If the ad wins and receives a click, charge $r+c+\frac{v}{p}$.
\end{itemize}

\section{Combining Per-View and Per-Click Charges for CPM Ads}
Let $v$ be the per-view charge and $c$ be the per-click charge. Let $b$ be the per-view bid.
\begin{itemize}
\item Enter value $a=b-v-cp$ as the per-view bid. 
\item If the ad wins, charge $r+v+cp$.
\end{itemize}

\section{The General Method}
This section outlines a general method to charge for impact on user experience in auctions. The previous results in this note are special cases of this general method. In addition to the price types discussed previously, this general method also applies to hybrid price types, where an advertiser offers to pay for multiple outcomes or events. (For example, an advertiser may offer some money to show their ad and an additional amount if the ad is clicked.)

For a set of events that may follow from selecting an ad, let variable $e_i$ be one if event $i$ occurs and zero if not. Examples of events include displaying the ad, a click on the ad, a conversion, etc. Let $b_i$ be the bid amount the advertiser is willing to pay if event $i$ occurs. Then the advertiser is willing to pay a total of:
$$ \sum_{i} b_i e_i. $$

Let $p_i$ be the probability of event $i$. (For the event that the ad is displayed, $p_i = 1$.) The advertiser’s offer has expected value
$$ \sum_{i} b_i p_i. $$ 
 
Let $c_i$ be the charge for impact on user experience when event $i$ occurs. Then a straightforward method to apply the charges is:
\begin{itemize}
\item For each $i$, enter $a_i = b_i - c_i$ as the per-event $i$ bid in the auction.
\item If the ad wins, then charge $\sum_{i} (r_i + c_i)e_i$, where $r_i$ is the auction-determined price for event $i$.
\end{itemize}
Then the expected payment is
$$ \sum_{i} (r_i + c_i)_pi.$$

This method adjusts each per-event bid to account for the charge for that event. But in many cases the advertiser is only willing to pay for some events, and the marketplace needs to charge for the impact on user experience caused by other events. In these cases, it is possible to achieve the same expected value for the advertiser’s adjusted offer in the auction and the same expected payment if the advertiser wins by shifting some charges among events.

Let $d_1, d_2, \dots$ be any charges such that
$$ \sum_{i} d_i p_i = \sum_{i} c_i p_i.$$
Then use the method:
\begin{itemize}
\item For each $i$, enter $a_i = b_i - d_i$ as the per-event $i$ bid in the auction.
\item If the ad wins, charge $\sum_{i} (r_i + d_i)e_i$, where $r_i$ is the auction-determined price for event $i$.
\end{itemize}
The offer adjusted by $d_1, d_2, \ldots$ has the same expected value as the offer adjusted by $c_1, c_2, \ldots$. Also, if the prices $r_i$ are the same under both adjustments, then
$$ \sum_{i} (r_i + d_i)p_i =  (r_i + c_i)p_i.$$
So there is the same expected payment if the advertiser wins.

If we want all per-event bids in the auction to be nonnegative, then we can impose the constraint on shifted charges:
$$ \forall i:d_i \leq b_i.$$
This constraint can be satisfied unless the expected impact on user experience for showing the ad is greater than the expected value of the advertiser’s offer, in which case the ad should not be shown at al, so its offer should not be entered into the auction.

\bibliographystyle{ACM-Reference-Format}
\bibliography{bax} 

\end{document}